\documentclass[superscriptaddress,twocolumn,aps,prl,showpacs]{revtex4-1}
\usepackage{epsfig}
\usepackage{graphicx}
\usepackage{dcolumn}
\usepackage{bm}
\usepackage{color}
\usepackage{ulem}
\usepackage{mathtools,amssymb}

\begin{document}

\title{Localization transition induced by learning in random searches}

\author{Andrea Falc\'on-Cort\'es}
\email{andreafalcon@estudiantes.fisica.unam.mx}
\affiliation{Instituto de F\'\i sica, Universidad Nacional Aut\'onoma de M\'exico, D.F. 04510, M\'exico}

\author{Denis Boyer}
\email{boyer@fisica.unam.mx}
\affiliation{Instituto de F\'\i sica, Universidad Nacional Aut\'onoma de M\'exico, D.F. 04510, M\'exico}

\author{Luca Giuggioli}
\email{Luca.Giuggioli@bristol.ac.uk}
\affiliation{Bristol Centre for Complexity Sciences, Department of Engineering Mathematics and School of Biological Sciences, University of Bristol, Bristol, BS8 1UB, UK}

\author{Satya N. Majumdar}
\email{majumdar@lptms.u-psud.fr}
\affiliation{Univ. Paris-Sud, CNRS, LPTMS, UMR 8626, Orsay F-91405, France}

\date{\today}

\begin{abstract}
We solve an adaptive search model where a random walker or L\'evy flight stochastically resets to previously visited sites on a $d$-dimensional lattice containing one trapping site. Due to reinforcement, a phase transition occurs when the resetting rate crosses a threshold above which
non-diffusive stationary states emerge, localized around the inhomogeneity. The threshold depends on the trapping strength and on the walker's return probability in the memoryless case.  
The transition belongs to the same class as the self-consistent theory of Anderson localization. These results show that similarly to many living organisms and unlike the well-studied Markovian walks, non-Markov movement processes can allow agents to learn about their environment and promise to bring adaptive solutions in search tasks.
\end{abstract}

\maketitle
Random searches have sparked enormous interest in recent years, as they find many applications in biology, physics and computer science \cite{kagan}. In typical settings, a target hidden in space has to be found by a searcher. The focus is commonly on first passage time statistics or the minimization of the mean searching time. Many theoretical approaches assume searchers lacking memory, which justifies Markovian dynamics such as the random walk (RW) \cite{vis11,benichou11,turchin98,colding08}. In contrast, allowing the searcher to gather and retain information about the environment may induce new adaptive behaviors, which can evolve in time as a result of experience. Namely, a learning process becomes possible. In engineering, a variety of robotic search tasks can be optimized by sampling more often spatial regions where targets are more likely to be present (see e.g. \cite{caisimon2014}). Likewise, animals seeking food, water or mates find it energetically convenient to revisit locations associated with successful searches \cite{morales14,barabasi101,wang14, boyer12,morales04}.
Exploiting regions rich in resources relies on two forms of memories, or a combination thereof: one that uses the environmental memory \cite{theraulazbonabeau99}, and one that uses the animal cognitive capabilities \cite{fagan13}. The former, which is well studied, is accomplished by depositing chemical substances, e.g. pheromone in ants, or physical marks to indicate the direct route to specific profitable locations \cite{bonabeauetal1999}. In this case a searcher would not require any memory, it could simply follow the scent trail once found. The latter form of resource exploitation, which inspires the present study, uses the actual ability of a searcher to remember previous positions and revisit them preferentially \cite{schacter1992}. 
Importantly, optimal uptake of available resources is often accomplished by a trade-off between frequently returning to known areas and randomly exploring uncharted ones \cite{vanmoorter,boyer10,bouchaud14}.

Path-dependent processes such as random walks with preferential revisits are mathematically challenging. For basic models on homogeneous lattices, the simple question whether asymptotic behaviors are diffusive or spatially localized is hard to tackle \cite{pemantle,abra10b,foster09}. Even less is understood on how spatial inhomogeneities may affect the properties of these processes, although an increasing number of biologically motivated models have been studied numerically \cite{borger08,gau06,boyer10,vincenot15}. Here, we solve analytically a model that combines random motion with a standard linear reinforcement scheme \cite{luce}, allowing to understand how spatial learning can emerge during a search. We find that above a critical threshold of memory use, non-Markovian effects can completely suppress diffusion at large times and localize the walk around a trapping site. This non-equilibrium phase transition is accompanied by a diverging length-scale and bears close similarities with the Anderson localization transition of waves in random media
\cite{abra10}.

Our approach is based on diffusion with resetting, a class of processes that have attracted a lot of 
attention for random search applications in the past few 
years~\cite{manrubia99,evans11,EM11,EM14,kus2014,eule,MSS15I,pal16,pal17,mendez16}. 
In those processes, standard diffusion is interrupted by stochastic resetting events that relocate the random walker back to its 
starting position (or some fixed position), leading to the emergence of non-equilibrium stationary states (NESS). 
Extensions including memory, where resetting can occur to any previously visited position, have also been 
studied \cite{boyer14,MSS15II,boyer17}. Here, let us consider a walker with position ${\mathbf{X}_t}$ on an infinite $d$-dimensional cubic lattice with unit spacing, where the time variable $t$ is discrete and the starting position is ${\mathbf X}_0={\mathbf x}_0$. The lattice contains one inhomogeneity, representing a water hole or a food patch, located at the origin. Depending on its position in space, the walker obeys two types of dynamics. ({\it i}) It follows a reinforced motion that combines diffusion and resetting to locations visited in the 
past \cite{gau06,boyer14,boyer16}, or ({\it ii}) it remains trapped at the origin for some time. More precisely, at each time step $t\rightarrow t+1$:
\begin{enumerate}
 \item[(a)] If the walker
is not at the inhomogeneity, with probability $1-q$ it selects a random displacement ${\boldsymbol \ell}_t$ drawn from a symmetric distribution $p(\boldsymbol{\ell})$ and ${\mathbf X}_{t+1}={\mathbf X}_{t}+\boldsymbol{\ell}_t$  ({\it RW motion}). With the complementary probability $q$ the walker resets to a site visited in the past, that is ${\mathbf X}_{t+1}={\mathbf X}_{t'}$ where $t'$ is a random integer uniformly chosen in the interval $[0,t]$. Therefore, the probability of choosing a particular site for relocation is proportional to the accumulated amount of time spent at that site
({\it linear reinforcement}).
 
 \item[(b)] If the walker occupies the inhomogeneity ($\mathbf{X}_t=\mathbf{0}$) it stays there at $t+1$ with probability $\gamma$ ({\it trapping or feeding}), or moves according to the rules (a) with probability  $1-\gamma$.
\end{enumerate}

\begin{figure}
  \centerline{\includegraphics*[width=0.4\textwidth]{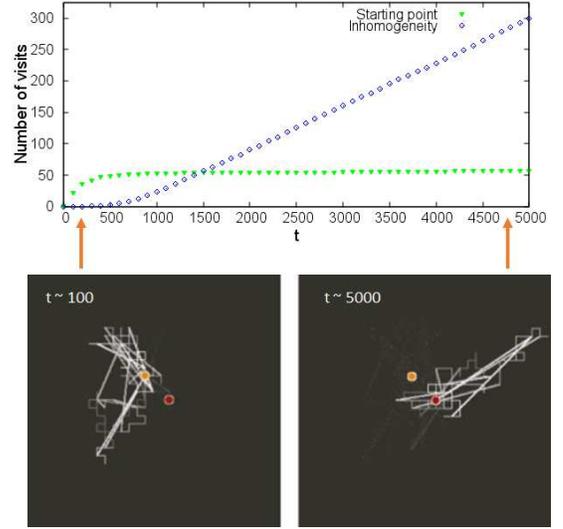}}
  \caption{(Color online) Average total number of visits to the starting site ${\mathbf x}_0=(-5,5)$ (green triangles) and to the inhomogeneity (blue diamonds) as function of time, for a $2d$ walker with $q=0.2$ $(\gamma=0.9)$. At early times, the walker slowly diffuses around ${\mathbf x}_0$ (orange-light disk in the insets). After the inhomogeneity (red-dark disk) has been found, it becomes steadily revisited, indicating spatial learning and localization.
  }
\label{fig2}
\end{figure}

By defining $\mbox{Prob}[\mathbf{X}_{t'}=\boldsymbol{n}\;\mbox{and}\;\mathbf{X}_t= 0]$ as the joint probability of being at $\boldsymbol{n}$ at time $t'$ and at the origin at time $t$, the above dynamics can be written as follows
\begin{eqnarray}\label{mastereq}
 P_{\boldsymbol{n}}(t+1)&=& (1-q)\sum_{\boldsymbol{\ell}} 
 p(\boldsymbol{\ell})(1-\gamma_{\boldsymbol{n}-\boldsymbol{\ell}})
 P_{\boldsymbol{n}-\boldsymbol{\ell}}(t) +\gamma_{\boldsymbol{n}}P_{\boldsymbol{n}}(t)\nonumber \\ 
 &+& \frac{q(1-\gamma)}{t+1}\sum\limits_{t'=0}^{t}\mbox{Prob}[\mathbf{X}_{t'}=\boldsymbol{n}
              \;\mbox{and}\;\mathbf{X}_t=\boldsymbol{0}]\nonumber\\
         &+& \frac{q}{t+1}\sum\limits_{t'=0}^{t}\mbox{Prob}[\mathbf{X}_{t'}=\boldsymbol{n}
              \;\mbox{and}\;\mathbf{X}_t\neq \boldsymbol{0}]
\end{eqnarray}
where $P_{\mathbf n}(t)=\mbox{Prob}[\mathbf{X}_t={\mathbf n}]$
and $\gamma_{\boldsymbol{n}}=\gamma\delta_{\boldsymbol{n},\boldsymbol{0}}$. 
The first two terms of the r.h.s. of Eq. (\ref{mastereq}) describe the random movement and trapping of the walker, respectively.
The last two terms of Eq. (\ref{mastereq}) account for the probability to reset to site $\boldsymbol{n}$ (if it has been visited at an earlier time $t'$) from a site that can be either the trapping site ${\mathbf 0}$ or another site. The term $1/(t+1)$ is the uniform probability distribution of the variable $t'$. 
Equation (\ref{mastereq}) describes a non-Markov process with infinite memory taking place in an inhomogeneous medium. In the absence of spatial heterogeneity ($\gamma=0$) the model exhibits unbounded (albeit very slow) diffusion for any memory strength or resetting probability ($0< q<1$): $\lim_{t\rightarrow \infty} P_{\mathbf n}(t)=0$ $\forall {\mathbf n}$ \cite{boyer14,boyer16}. When $\gamma\ne0$, each visit at the origin tends to last longer than at any other site and we ask whether reinforcement can suppress diffusion altogether and attract the dynamics toward a NESS, namely $P_{\mathbf n}\equiv\lim_{t\rightarrow \infty} P_{\mathbf n}(t)\neq 0$, centered around ${\mathbf 0}$ and independent of the walker initial position ${\mathbf x}_0$. When the NESS state is reached we say that the walker has localized as a result of adaptation by learning (see Fig. \ref{fig2}).

To gain valid insights on the dynamics of the learning searcher, we  consider an approximate version of the model. We make the assumption (to be checked later on) that at large times ${\mathbf X}_{t'}$ and ${\mathbf X}_t$ become uncorrelated:
\begin{equation}\label{limits1}
 \mbox{Prob}[\mathbf{X}_{t'}= \boldsymbol{n}
 \;\mbox{and}\;\mathbf{X}_t\neq \boldsymbol{0}] \simeq 
 P_{\boldsymbol{n}}(t')[1-P_{\boldsymbol{0}}(t)].
\end{equation}
Similarly, $\mbox{Prob}[\mathbf{X}_{t'}=\boldsymbol{n}\;\mbox{and}\;
\mathbf{X}_t=\boldsymbol{0}] \simeq P_{\boldsymbol{n}}(t')P_{\boldsymbol{0}}(t)$. Replacing these expressions in Eq. (\ref{mastereq}) and substituting $P_{\mathbf n}(t)$ and $P_{\mathbf n}(t')$ by $P_\mathbf n$ in the limit $(t,t')\rightarrow\infty$, we obtain an equation satisfied by the NESS:
\begin{eqnarray}\label{levyrecursive}
 P_{\mathbf n} &=& (1-q)\sum_{\boldsymbol \ell} p({\boldsymbol \ell})
P_{{\mathbf n}-{\boldsymbol \ell}}+qP_{\mathbf n}(1-\gamma P_{0})\\
     &+& \gamma P_0[\delta_{{\mathbf n},{\mathbf 0}}-(1-q)p({\mathbf n})], \nonumber
\end{eqnarray}
and valid for any number of dimensions. Note that if the summands have a finite limit, the last two terms of Eq. (\ref{mastereq}) do not vanish at large $t$. We note $P_0\equiv P_{{\boldsymbol 0}}$ as the asymptotic probability of occupying the inhomogeneity, a quantity yet to be determined.
We introduce the discrete Fourier transform 
$\tilde{f}(\mathbf{k})\equiv\sum_{\mathbf n} f_{\mathbf n} e^{-i{\mathbf k}\cdot\mathbf{n}}$ \cite{feller08}. Transforming Eq. (\ref{levyrecursive}) yields:
\begin{equation}\label{transPn}
 \tilde{P}({\mathbf k})=\frac{\gamma P_0[1-(1-q)\tilde{p}({\mathbf k})]}{(1-q)[1-\tilde{p}({\mathbf k})]+q\gamma P_0}.
\end{equation}
The constant $P_0$ is determined self-consistently from the inverse transform of (\ref{transPn}) evaluated at ${\mathbf n}={\mathbf 0}$: $P_0=(2\pi)^{-d}\int_{\cal B}d^d{\mathbf k}\ \tilde{P}({\mathbf k})$, where ${\cal B}$ is the first Brillouin zone: $-\pi<k_i<\pi$ for $i=1,\ldots,d$. After re-arrangements, any solution $P_0\ne0$ obeys the transcendental equation:
\begin{equation}\label{levyP0}
 \frac{1}{(2\pi)^d}\int_{\cal B} \frac{d^d{\mathbf k}}{(1-q)[1-\tilde{p}({\mathbf k})]+q\gamma P_0} = \frac{1-\gamma}{q\gamma(1-\gamma P_0)}.
\end{equation}
Fixing $\gamma>0$, the model exhibits a phase transition if there exists a critical $q_c\in(0,1)$ such that Eq. (\ref{levyP0}) does not have any root for $q<q_c$ (in such case, only the trivial solution $P_0=0$ exists). Hence, setting $P_0=0$ in Eq. (\ref{levyP0}) gives the threshold $q_c$:
\begin{equation}\label{qc}
 q_c=\frac{(1-\gamma)P_{no-return}}{\gamma+(1-\gamma)P_{no-return}},
\end{equation}
where $P_{no-return}=(2\pi)^{d}[\int_{\cal B}d^d\mathbf{k}\ \frac{1}{1-\tilde{p}(\mathbf{k})}]^{-1}$ is the well-known probability for the Markovian random walk on the infinite lattice to never come back to its starting site \cite{feller08}. This is a remarkably simple result, reminiscent of the phenomenology of the Anderson transition: a delocalization/localization transition can exist at some $q_c>0$ (or $\gamma_c>0$ if $q$ is held fixed) if $P_{no-return}>0$, {\it i.e.}, if the process with $q=0$ (no resetting) may never return to its starting site, such as the nearest neighbor (n.n.) RW in $d\ge3$. Conversely, recurrent processes like RWs in $1d$ and $2d$ have $P_{no-return}=0$ and thus admit localized solutions when memory is switched on to any strength $q>0$. We emphasize that for a pure RW ($q=0$), a single impurity of any finite strength $\gamma<1$ is not enough to localize the walker. We also mention that 
the case $q=1$ is pathological since the walker stays immobile and thus cannot find the origin, unless ${\mathbf x}_0={\mathbf 0}$.

Before discussing properties of the critical point, we exactly solve Eq. (\ref{transPn}) in the particular 1\textit{d} case with n.n. hopping, where $\tilde{p}(k)=\cos(k)$. By Fourier inversion we find (see \cite{Supp_Mat}):
\begin{equation}\label{Pn}
 P_n=\gamma P_0\delta_{n,0}+(1-\gamma)P_0a^{-|n|},
\end{equation}
with $a=1+\frac{\gamma q P_0}{1-q}+\sqrt{\frac{\gamma qP_0 }{1-q}\left(2+\frac{\gamma qP_0 }{1-q}\right)}$, and
\begin{eqnarray}\label{P0}
 P_0 &=& \frac{-(1-q)(1-\gamma)^2-q\gamma^2}{q\gamma(1-2\gamma)}\\ \nonumber
     &+& \frac{\sqrt{[(1-q)(1-\gamma)^2+q\gamma^2]^2+(q\gamma)^2(1-2\gamma)}}{q\gamma(1-2\gamma)},
\end{eqnarray}
for $\gamma\ne1/2$. When $\gamma=1/2$, one simply obtains $P_0=q$.

\begin{figure}
  \centerline{\includegraphics*[width=0.4\textwidth]{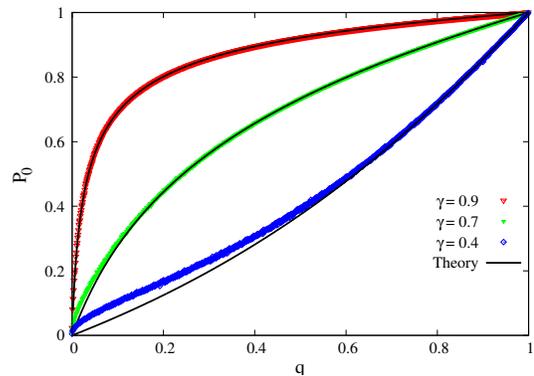}}
  \caption{(Color online) $P_0$ as a function of $q$ in $1d$ with n.n. hopping. Solid lines are given by Eq. (\ref{P0}) and symbols by Monte Carlo simulations of the rules (a)-(b) (at $t=10^5$).}
\label{fig1}
\end{figure}

Eq. (\ref{P0}) shows that $P_0>0$ for any $\gamma>0$ and $q>0$: as previously announced {\it localized solutions always exist in 1d} for any memory and inhomogeneity strengths. The probability of presence decays exponentially with the distance to the origin. 
Fig. \ref{fig1} displays $P_0$ as a function of $q$ for different $\gamma$, as given by Eq. (\ref{P0}). Instead of solving Eq. (\ref{mastereq}) numerically, which is difficult, we have performed Monte Carlo simulations of rules (a)-(b). The very good agreement obtained suggests that our de-correlation approximation might be exact at large times. The discrepancy observed at small $q$ and $\gamma$ is attributed to the fact that the asymptotic time regime is very long to reach in simulations: For $\gamma=0$, diffusion is logarithmic and $P_0$ tends to $0$ as $1/\sqrt{\ln\;t}$ \cite{boyer14}. To accelerate the convergence to the NESS, we set ${\mathbf x}_0=0$ in all the simulations presented here \cite{notex0}. 

\begin{figure}
  \centerline{\includegraphics*[width=0.55\textwidth]{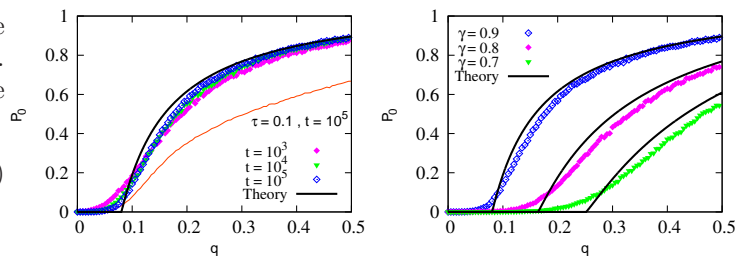}}
  \caption{(Color online) Phase transition in $1d$ for L\'evy flights with $\mu=1/2$. Left: The thick (black) solid lines are given by Eq. (\ref{levyP0}) for the $t=\infty$ limit, and the symbols by simulations up to different times $t$ ($\gamma=0.9$). The parameter $\tau$ is explained in the text, end of p. 4. Right: $P_0$ for $\gamma=0.9$, $0.8$ and $0.7$, where theory predicts $q_c\approx 0.0803$, $0.1642$ and $0.2519$, respectively ($t=10^5$ in simulations).}
\label{fig3}
\end{figure}

We next study some of the $1d$ cases in which $P_{no-return}>0$, implying a phase transition at a non-zero $q_c$ (see \cite{Supp_Mat}). For this purpose we consider a symmetric L\'evy flight (LF) whereby
\begin{equation}\label{levydis}
 p(\ell)=C/|\ell|^{1+\mu},\;\;\;\ell=\pm 1,\pm 2,\pm 3...,
\end{equation}
with index $0<\mu\leq 1$ \cite{samo94,weiss94}. Figure \ref{fig3} displays $P_0$ as a function of $q$ (for $\mu=1/2$ and several $\gamma$), as given by a numerical solution of Eq. (\ref{levyP0}) and by Monte Carlo simulations of the walker dynamics. A very good agreement is obtained. As expected, at larger simulation times the variations of $P_0$ become steeper around the critical point (left panel). 

\begin{figure}
  \centerline{\includegraphics*[width=0.4\textwidth]{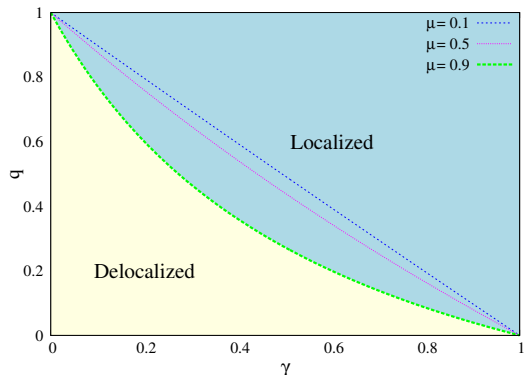}}
  \caption{(Color online) Phase diagram in $1d$ for various L\'evy indexes.}
\label{fig4}
\end{figure}

From Eq. (\ref{qc}), we draw the $1d$ phase diagram in the $(\gamma,q)$-plane in Figure (\ref{fig4}). The thick (green) dashed curve represents the line of critical points for $\mu=0.9$. Processes characterized by $P_{no-return}\rightarrow 1$, can be obtained either by taking the limit $\mu\rightarrow0$ in $1d$ or the limit $d\rightarrow\infty$ (where any process is expected to become highly transient). These cases correspond to the diagonal $q_c=1-\gamma$.

The general critical behavior of $P_0$ in any $d$, for generic LFs including the standard RW case,
is obtained from a Taylor expansion of Eq. (\ref{levyP0}) near $q_c$. 
Since the small $\mathbf{k}$ regime dominates, one uses the expansion 
$1-\tilde{p}({\mathbf k})\simeq K_{\mu}|\mathbf{k}|^{\mu}$ (for LFs) or $1-\tilde{p}({\mathbf k})\simeq D_0|\mathbf{k}|^2$ 
(for normal RWs, if $p({\boldsymbol \ell})$ has finite variance) \cite{weiss94}. By analyzing the integral in $P_{no-return}$, it is easy 
to see~\cite{Supp_Mat} that for $d>d_c=\mu$ the walk is transient, implying $P_{no-return}>0$ and consequently from Eq. (\ref{qc}), $q_c>0$.
In contrast, for $d<d_c= \mu$, the walk is recurrent, implying $P_{no-return}=0$ and hence $q_c=0$. Near $q=q_c$, we find~\cite{Supp_Mat}
\begin{equation}\label{nearqc}
 P_0\sim (q-q_c)^{\beta},
\end{equation}
in all cases, where $\beta= 1$ for $d>2\mu$, $\beta= \mu/(d-\mu)$ for $\mu<d<2\mu$, while $\beta= d/(\mu-d)$ for $d<\mu$ (in this last case $q_c=0$). Normal RW's correspond to the special case $\mu=2$: if $d<2$, the transition takes place at $q_c=0$ and $\beta=d/(2-d)$; if $2<d<4$, then $\beta=2/(d-2)$; if $d>4$, then $\beta=1$. 
These exponents and critical dimensions are actually identical to those of the self-consistent theory (SCT) of 
Anderson localization \cite{vollhardt,vollhardt2,abra10}. In this problem the diffusion coefficient describing wave transport in disordered media obeys a self-consistent relation similar to Eq. (\ref{levyP0}).

Once $P_0$ is known, the large $|{\mathbf n}|$ behavior of $P_{\mathbf n}$ is readily obtained. At small ${\mathbf k}$, Eq. (\ref{transPn}) gives, for the normal RW case and $q>q_c$:
\begin{equation}\label{limitdis}
 \tilde{P}({\mathbf k})\simeq\frac{q^*}{D|{\mathbf k}|^2+q^*},\;\;\;{\rm with}\;\;q^*=q\gamma P_0,
\end{equation}
and $D=(1-q)D_0$ a rescaled diffusion constant. Up to a prefactor, this form coincides with the correlation function of the Gaussian model of second-order phase transitions with scalar order parameter, $\tilde{C}({\mathbf k})\propto 1/(|{\mathbf k}|^2+\xi^{-2})$, which stems, like the SCT of Anderson localization, from a one-loop approximation. Hence $\xi=(D/q^*)^{1/2}$ is the localization length. The inverse transform of (\ref{limitdis}) decays exponentially at large $|{\mathbf n}|$, see \cite{evans11} or \cite{gold} for its precise form in all $d$. From (\ref{nearqc})-(\ref{limitdis}), one deduces that $\xi$ always diverges as $(q-q_c)^{-\nu}$ near $q_c$, with $\nu=\beta/2$. Therefore $\nu=d/(4-2d)$ if $d<2$; $\nu=1/(d-2)$ if $2<d<4$; and $\nu=1/2$ if $d>4$. These exponents are again those of the SCT of Anderson localization \cite{vollhardt}.

Remarkably, the distribution (\ref{limitdis}) also has the same expression than the NESS generated by diffusion with stochastic resetting to the unique site ${\mathbf 0}$ at rate $q^*$ \cite{evans11,kus15}. Therefore, thanks to learning, the walker effectively behaves at large times like a memoryless walker that resets to the inhomogeneity {\it only}. The selection of the resetting point is an emergent property, and not imposed like in \cite{evans11,kus15}. The effective resetting rate $q^*$ is $\propto P_0$ and thus vanishes at $q=q_c$, where the walker is no longer able to adapt to its environment.

\begin{figure}
  \centerline{\includegraphics*[width=0.45\textwidth]{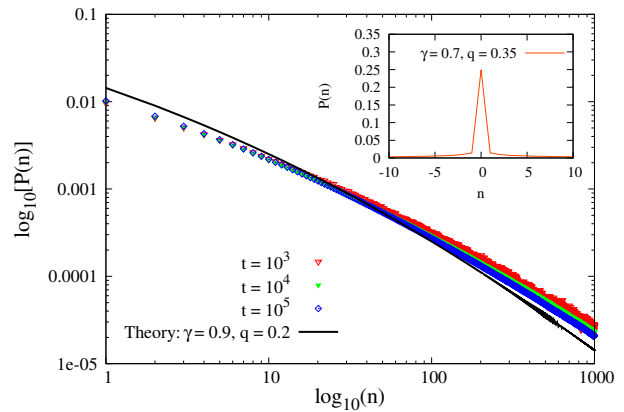}}
  \caption{Asymptotic probability of presence in $1d$ for L\'evy flights with $\mu=1/2$. The solid line is the inverse Fourier transform of Eq. (\ref{transPn}) and symbols represent simulations.}
\label{fig5}
\end{figure}

In the case of L\'evy flights, $D|{\mathbf k}|^2$ is replaced by $K|{\mathbf k}|^{\mu}$ in (\ref{limitdis}): this expression also coincides with the NESS for a LF with resetting at the origin \cite{kus2014,kus15}. In $1d$ and for $\mu\in(0,2)$, the inversion gives the Linnik distribution \cite{kotz95}:
\begin{equation}\label{PnLap}
 P_n\simeq\left\{\frac{K}{q^{*}\pi} \sin\left(\frac{\pi\mu}{2}\right)\Gamma(\mu +1)\right\}
 |n|^{-1-\mu}+R(n)
\end{equation}
with $|R(n)|<\{(K/q^{*})^2\Gamma(\mu +1)/\pi\sin(\pi\mu/2)\} 
|n|^{-1-2\mu}$.
The walker is thus power-law localized, with exponent $-(1+\mu)$ at large $|n|$. Fig.(\ref{fig5}) shows the good agreement between $P_n$ obtained from numerical inversion and simulations at $\mu=0.5$.

The robustness of the localization phenomenon can be probed by incorporating resource depletion and refreshing in rule (b). Let us assume that the inhomogeneity becomes empty each time the walker leaves it ($\gamma$ set to $0$), and then recovers ($0\rightarrow\gamma$) at a later time with rate $\tau$. Fig. \ref{fig3}-left displays a simulation curve of $P_0$ for $\tau=0.1$, whose shape is similar to that of the base model. 


In summary, we have demonstrated with a solvable model that random walkers with resetting and memory are able to learn by reinforcement and adapt to features of their environments. Adaptation is revealed through a localization phase transition which emerges around a trapping site. 
The localized walker asymptotically behaves as if it reset to the trapping site solely, with an effective resetting rate that vanishes at criticality. 
Apart from applications in cognition and ecology, the results presented here could motivate applications for pattern recognition, the tracking of mobile objects, or for developing algorithms that solve difficult optimization problems. Our study also establishes a long sought formal analogy between the localization of path-dependent random walks and that of waves in disordered media \cite{abra10b}. Despite the radically different physical nature, non-Markovian stochastic processes with resetting may prove useful for studying Anderson transitions. 

DB and LG acknowledge support from the Max Planck Institute for the Physics of Complex Systems and its Advanced Study Group \lq\lq Anomalous diffusion in foraging". DB acknowledges support from DGAPA-PAPIIT Grant IN105015, and LG from EPSRC grant EP/I013717/1. We thank A. Aldana, M. Aldana, J. Cardy, F. Leyvraz, and G. G. Naumis for fruitful discussions. 


\newpage

\begin{widetext}

\vspace{1.5cm}
\begin{center}
\hskip 0cm {\Large{Supplemental Material}}
\end{center}

\vspace{0.5cm}

\section{I. No-return Probability for L\'evy flights: Recurrent vs. Transient behavior}
\label{sec:Escape_Levy}

Consider a $d$-dimensional Euclidean lattice. A random walker moves on the sites of this lattice with random jumps at each time step.
The jump lengths are independent and identically distributed (i.i.d) random variables drawn from a normalized distribution
$p({\boldsymbol \ell})$. The walker starts at some arbitrary site $({\boldsymbol x_0}$ at time $t=0$). Then the probability
of no return to the initial site is given by the well known formula
\begin{equation}
P_{no-return}=\frac{1}{\int_{\cal B}\frac{d^d\mathbf{k}}{(2\pi)^d}\, \frac{1}{1-\tilde{p}(\mathbf{k})} }
\label{no_return_SM1}
\end{equation}
where $\tilde{p}(\mathbf{k})$ is the Fourier transform of the jump distribution 
\begin{equation}
\tilde{p}(\mathbf{k})= \sum_{\mathbf \ell} p({\boldsymbol \ell}) e^{-i{\mathbf k}\cdot\mathbf{\ell}}\, .
\label{ft_pl_SM2}
\end{equation}
Thus $P_{no-return}$ in Eq. (\ref{no_return_SM1}) is nonzero or zero depending on whether the integral in the denominator
is finite or divergent. The divergence of this integral depends on the small $|\mathbf{k}|$ behavior of
$\tilde{p}(\mathbf{k})$. In general, for L\'evy flights, the small $k$ behavior is given by
\begin{equation}
\tilde{p}(\mathbf{k})\simeq 1- K_{\mu}|\mathbf{k}|^{\mu}
\label{LF1_SM3}
\end{equation}
where the L\'evy index $0<\mu\le 2$. For $\mu<2$, the second moment of the jump distribution is divergent, while for $\mu=2$, the
second moment is finite. Hence, standard Euclidean random walks with nearest neighbour jumps correspond to $\mu=2$,
with $\tilde{p}({\mathbf k})=1-  D_0|\mathbf{k}|^2$. From now on, we will consider the general $0<\mu\le 2$ case, and
it will include the $\mu=2$ case corresponding to  
standard nearest neighbour random walks. Substituting the small $k$ behavior in the integral in the denominator of
Eq. (\ref{no_return_SM1}), it is evident that this integral diverges if $d<\mu$ and is finite if $d>\mu$. Thus, for
$d<\mu$, $P_{no-return}=0$, while it is non zero for $d>\mu$. Thus, for L\'evy flights with index $\mu$ ($0<\mu\le 2$), the
critical dimension is $d_c=\mu$ that separates the recurrent ($d<\mu$) behavior from the transient ($d>\mu$) behavior. For ordinary
random walks ($\mu=2$), $d_c=2$. 

\section{II. Critical behavior of the order parameter $P_0$}
\label{sec:critical_P0}

We first consider the critical value $q_c$ (for fixed $\gamma$) that separates the delocalised phase with $P_0=0$ for $q<q_c$
and the localised phase with $P_0>0$ for $q>q_c$. In the main text, we have shown that the value of $q_c$ is given by the formula
\begin{equation}
q_c=\frac{(1-\gamma)P_{no-return}}{\gamma+(1-\gamma)P_{no-return}}
\label{qc_SM4}
\end{equation}
where $P_{no-return}$ is given in Eq. (\ref{no_return_SM1}). So, clearly, for L\'evy flights with index $0<\mu\le 2$ (including
standard random walks corresponding to $\mu=2$), using results on $P_{no-retun}$ from the previous Section \ref{sec:Escape_Levy},
we have 
\begin{eqnarray}
q_c &= & \frac{(1-\gamma)P_{no-return}}{\gamma+(1-\gamma)P_{no-return}}>0 \quad\,\, {\rm for}\quad d>\mu \label{qcfinite} \\
& =& 0  \hskip 4cm \quad {\rm for}\quad d<\mu \, . \label{qc0}
\end{eqnarray}

We now consider how $P_0$ increases from its value $0$ as $q$ increases above $q_c$. We want to show here that in general,
as $q\to q_c^{+}$, 
\begin{equation}
P_0 \sim (q-q_c)^{\beta}
\label{beta_SM5}
\end{equation}
where the exponent $\beta$ depends continuously on $\mu$ and $d$ in the $\mu-d$ plane. We will show below that
\begin{eqnarray}
\beta= 
\begin{cases}
 1 \quad\quad\,\,\,\, {\rm for} \quad d>2\mu \label{upper} \\
\frac{\mu}{d-\mu} \quad\,\, {\rm for} \quad \mu<d<2\mu \label{middle} \\
\frac{d}{\mu-d}\quad\,\, {\rm for} \quad  d<\mu \label{lower}
\end{cases}
\end{eqnarray}
where, we recall, that in the last case ($d<\mu$), $q_c=0$.

To derive this result for $\beta$, we start from the equation in the main text that determines $P_0$ for any given $q$, namely
\begin{equation}
\frac{1}{(2\pi)^d}\int_{\cal B} \frac{d^d{\mathbf k}}{(1-q)[1-\tilde{p}({\mathbf k})]+q\gamma P_0} = 
\frac{1-\gamma}{q\gamma(1-\gamma P_0)} \, .
\label{P0_SM6}
\end{equation}
Of course, at $q=q_c$, $P_0=0$ and this gives us 
\begin{equation}
\frac{1}{(2\pi)^d}\int_{\cal B} \frac{d^d{\mathbf k}}{(1-q_c)[1-\tilde{p}({\mathbf k})]}= \frac{1-\gamma}{q_c\,\gamma}\, ,
\label{qc1_SM7}
\end{equation}
which indeed leads to the expression for $q_c$ in Eq. (\ref{qc_SM4}). 

We are now ready to see how $P_0$ increases from $0$ as $q$ increases above $q_c$. For this we consider two cases 
separtaely.

\vskip 0.4cm

\noindent {\bf Case I: ${\mathbf q_c>0}$.} As we have seen before, this corresponds to the transient regime where 
$P_{no-return}>0$. For L\'evy flights, this means $d>d_c=\mu$. To proceed,
we first subtract Eq. (\ref{P0_SM6}) from Eq. (\ref{qc1_SM7}) which gives
\begin{equation}
\int_{\cal B} \frac{d^d{\mathbf k}}{(2\pi)^d}\, \frac{[q\gamma \delta-
(q-q_c)(1-\tilde{p}({\mathbf k}))]}{(1-\tilde{p}({\mathbf k}))[(1-q)(1-\tilde{p}({\mathbf k}))+ q\gamma P_0]}
= \frac{(1-q_c)(1-\gamma)(q-q_c-q\gamma P_0)}{qq_c \gamma (1-\gamma P_0)}\, .
\label{lin_SM8}
\end{equation}
We then set $q=q_c+\epsilon$ with $\epsilon\to 0$
and $P_0=\delta$ with $\delta\to 0$. Our goal is to find how $\delta$ scales with $\epsilon$ to leading order in
small $\epsilon$. In this limit, the leading contribution to the integral on the left hand side (lhs) of 
Eq. (\ref{lin_SM8}) comes from the small $k$ region, where we can replace $\tilde{p}({\mathbf k})$ by
Eq. (\ref{LF1_SM3}). Keeping only the leading order terms and simplifying, we obtain
\begin{equation}
\delta I(\delta) + O(\delta) = A \epsilon
\label{lin1_SM9}
\end{equation}
where $A= (1-\gamma)(1-q_c) K_\mu^2/(\gamma^2 q_c^3)$ is just a constant and $I(\delta)$ is the integral
\begin{equation}
I(\delta)= \int_{\cal B} \frac{d^d{\mathbf k}}{(2\pi)^d}\, \frac{1}{|{\mathbf k}|^{\mu}[|{\mathbf k}|^{\mu}+ b \delta]}
\label{Id_SM10}
\end{equation}
where $b= q_c\gamma/(K_\mu(1-q_c))$ is a constant. We now need to analyse the integral $I(\delta)$ as $\delta\to 0$.
There are again two cases: (1) $d>2\mu$ and (2) $\mu<d<2\mu$. We consider them separately.

\begin{enumerate}

\item $d>2\mu$: In this case, if we put $\delta=0$ in $I(\delta)$ in Eq. (\ref{Id_SM10}), 
the integral converges as $k\to 0$, making $I(0)$ 
finite. Hence, from Eq. (\ref{lin1_SM9}), we get
\begin{equation}
\delta \sim \epsilon\quad {\rm implying}\quad \beta=1 \quad {\rm for}\quad d>2\mu\, .
\label{beta_upper_SM11}
\end{equation}

\item{$\mu<d<2\mu$}: In this case, the integral $I(0)$ in Eq. (\ref{Id_SM10}) is divergent. 
Hence, to extract the leading singularity, we rescale $k\to \delta^{1/\mu} y$ in Eq. (\ref{Id_SM10}).
\begin{equation}
I(\delta) \sim \delta^{\frac{d}{\mu}-2} \int_0^{\infty} \frac{dy\, y^{d-1-\mu}}{y^\mu+b}\,.
\label{Id1_SM12}
\end{equation}
Note that the integral in Eq. (\ref{Id1_SM12}) is convergent in both limits $y\to 0$ and $y\to \infty$, as long as
$\mu<d<2\mu$. Hence, substituting Eq. (\ref{Id1_SM12}) in Eq. (\ref{lin1_SM9}) we get, to leading order
\begin{equation}
\delta \sim \epsilon^{\frac{\mu}{d-\mu}}\quad {\rm implying}\quad \beta= \frac{\mu}{d-\mu} \quad {\rm for}\quad \mu<d<2\mu \, .
\label{beta_middle_SM13}
\end{equation}

\end{enumerate} 

\vskip 0.4cm

\noindent {\bf Case II: ${\mathbf q_c=0}$.} This case corresponds to the recurrent case when $P_{no-return}=0$, making
$q_c=0$. As discussed before, for L\'evy flights with index $0<\mu\le 2$, this happens when $d<d_c=\mu$.
In this case we analyse directly Eq. (\ref{P0_SM6}) by substituting $q=\epsilon$ and $P_0=\delta$. Again, keeping
only the small ${\mathbf k}$ contribution to the integral, we get to leading order
\begin{equation}
 \int_{\cal B} \frac{d^d{\mathbf k}}{(2\pi)^d} \frac{1}{|{\mathbf k}|^\mu+ \epsilon \delta}\sim \frac{1}{\epsilon}\,
\label{lin2_SM14}
\end{equation}
Rescaling $k= (\epsilon \delta)^{1/\mu} y$ gives
\begin{equation}
(\epsilon \delta)^{\frac{d}{\mu}-1} \int_0^{\infty} \frac{dy\, y^{d-1}}{y^{\mu}+1} \sim \frac{1}{\epsilon}\, .
\label{lin3_SM15}
\end{equation}
Note that the integral in Eq. (\ref{lin3_SM15}) is convergent in both limit $y\to 0$ and $y\to \infty$ for
$0<d<\mu$. Hence, Eq. (\ref{lin3_SM15}) then gives
\begin{equation}
\delta \sim {\epsilon}^{\frac{d}{\mu-d}} 
\quad {\rm implying}\quad \beta= \frac{\mu}{\mu-d} \quad {\rm for}\quad 0<d<\mu\, .
\label{beta_lower_SM16}
\end{equation}
This completes the derivation of the result for the exponent $\beta$ given in Eqs. (\ref{upper}), (\ref{middle})
and (\ref{lower}).

\section{III. Localization of the $1d$ random walk with nearest neighbors jumps}
\label{sec:1dnn}

We derive here an analytical expression for the stationary distribution $P_n$. We consider the particular case of the random walk with nearest neighbor jumps in one dimension, where the step distribution is given by 
$p(l)=\frac{1}{2}[\delta_{l,1}+\delta_{l,-1}]$. The Fourier transform of $p(l)$ is $\tilde{p}(k)=\cos k$. In this case,
the expression given by Eq. (4) of the main text for the Fourier transform of $P_n$ becomes
\begin{equation}\label{pk1d}
\tilde{P}(k)=\frac{\gamma P_0[1-(1-q)\cos k]}{(1-q)(1-\cos k)+q\gamma P_0}=\gamma P_0
+\frac{q\gamma P_0(1-\gamma P_0)}{(1-q)(1-\cos k)+q\gamma P_0}.
\end{equation}
The form of the steady-state probability can be derived by inverse Fourier transforming. Using the fact that for $a^2>1$ 
\cite{gradshteynryzhik2015}:   
\begin{equation}\label{grad}
\frac{1}{2\pi}\int_{-\pi}^{\pi}dk\frac{\cos(kn)}{1+a^2-2a\cos k }=\frac{1}{(a^2-1)a^{|n|}}, 
\end{equation}
we write the denominator $(1-q)(1-\cos k)+q\gamma P_0$ under the form $b(1+a^2-2a\cos k)$. By identification, we have:
\begin{eqnarray}
2ab&=&1-q \label{a} \\
b(1+a^2)&=&1-q(1-\gamma P_0)\label{b}
\end{eqnarray}
which yields 
\begin{equation}\label{sola}
a=1+\frac{\gamma q P_0}{1-q}+\sqrt{\frac{\gamma qP_0 }{1-q}\left(2+\frac{\gamma qP_0 }{1-q}\right)}.
\end{equation}
Using Eq. (\ref{grad}) and (\ref{a}), the inversion of Eq. (\ref{pk1d}) gives:
\begin{equation}
P_n=\gamma P_0\delta_{n,0}+\frac{q\gamma P_0(1-\gamma P_0)}{1-q}\frac{2a}{(a^2-1)a^{|n|}}.
\label{eqn:exactPn1d}
\end{equation}
By evaluating  Eq. (\ref{eqn:exactPn1d}) at $n=0$, the above expression can be rewritten in compact form:
\begin{equation}\label{Pn}
 P_n= \gamma P_0\delta_{n,0} +(1-\gamma)P_0a^{-|n|},
\end{equation}
which is one of the main result of this section. We are only left with the determination of
$P_0$, the asymptotic probability of occupying the inhomogeneity. For this purpose, we evaluate once more Eq. (\ref{eqn:exactPn1d}) at $n=0$, obtaining:
\begin{equation}\label{aP0}
2q\gamma(1-\gamma P_0)=(1-\gamma)(1-q)(a-a^{-1}).
\end{equation}  
Inserting the expression of $a$ given by Eq. (\ref{sola}) into Eq. (\ref{aP0}) gives a quadratic equation for $P_0$ whose only positive root is 
\begin{equation}\label{P0}
 P_0 = \frac{-(1-q)(1-\gamma)^2-q\gamma^2}{q\gamma(1-2\gamma)}
     + \frac{\sqrt{[(1-q)(1-\gamma)^2+q\gamma^2]^2+(q\gamma)^2(1-2\gamma)}}{q\gamma(1-2\gamma)},
\end{equation}
for $\gamma\ne1/2$. When $\gamma=1/2$, the solution is simply $P_0=q$.

%
%
\end{widetext}

\end{document}